\documentclass[twocolumn,english,final,showpacs,prl]{revtex4}
\usepackage[T1]{fontenc}
\usepackage[latin9]{inputenc}
\usepackage{amsmath}
\usepackage{graphicx}
\usepackage{amssymb}

\makeatletter
\@ifundefined{textcolor}{}
{%
 \definecolor{BLACK}{gray}{0}
 \definecolor{WHITE}{gray}{1}
 \definecolor{RED}{rgb}{1,0,0}
 \definecolor{GREEN}{rgb}{0,1,0}
 \definecolor{BLUE}{rgb}{0,0,1}
 \definecolor{CYAN}{cmyk}{1,0,0,0}
 \definecolor{MAGENTA}{cmyk}{0,1,0,0}
 \definecolor{YELLOW}{cmyk}{0,0,1,0}
 }

\usepackage{textcomp}
\usepackage{babel}

\usepackage{amsfonts}

\providecommand{\U}[1]{\protect\rule{.1in}{.1in}}
\makeatletter
\providecommand{\LyX}{L\kern-.1667em\lower.25em\hbox{Y}\kern-.125emX\@}
\@ifundefined{textcolor}{}
{
\definecolor{BLACK}{gray}{0}
\definecolor{WHITE}{gray}{1}
\definecolor{RED}{rgb}{1,0,0}
\definecolor{GREEN}{rgb}{0,1,0}
\definecolor{BLUE}{rgb}{0,0,1}
\definecolor{CYAN}{cmyk}{1,0,0,0}
\definecolor{MAGENTA}{cmyk}{0,1,0,0}
\definecolor{YELLOW}{cmyk}{0,0,1,0}
}
\makeatother
\makeatother

\makeatother

\usepackage{babel}

\makeatother

\usepackage{babel}

\begin{document}

\preprint{This line only printed with preprint option}

\title{NMR quantum simulation of localization effects induced by decoherence}

\author{Gonzalo A. \'{A}lvarez}

\email{galvarez@e3.physik.uni-dortmund.de}

\affiliation{Fakult\"{a}t Physik, Universit\"{a}t Dortmund, Otto-Hahn-Strasse 4, D-44221
Dortmund, Germany.}

\author{Dieter Suter}

\email{Dieter.Suter@tu-dortmund.de}

\affiliation{Fakult\"{a}t Physik, Universit\"{a}t Dortmund, Otto-Hahn-Strasse 4, D-44221
Dortmund, Germany.}


\begin{abstract}
The loss of coherence in quantum mechanical superposition states limits
the time for which quantum information remains useful. Similarly,
it limits the distance over which quantum information can be transmitted,
resembling Anderson localization, where disorder causes quantum mechanical
states to become localized. Here, we investigate in a nuclear spin-based
quantum simulator, the localization of the size of spin clusters that
are generated by a Hamiltonian driving the transmission of information,
while a variable-strength perturbation counteracts the spreading.
We find that the system reaches a dynamic equilibrium size, which
decreases with the square of the perturbation strength. 
\end{abstract}
\maketitle
\textit{Introduction.---} Quantum information processing has the potential
of solving computational problems for which no efficient solution
exists on classical computers \cite{Shor1994,DiVincenzo1995,Nielsen00}.
Transfer and exchange of quantum information and quantum entanglement
can be used for secure transmission of information \cite{Cirac1997,Bose2003,Paternostro2005,Yang2009}.
Realization of this potential for practical applications requires
precise control of large quantum registers. However as the number
of qubits increases, the quantum mechanical superposition states of
the system become more fragile \cite{Suter04}. This degradation of
quantum superpositions, called decoherence \cite{Zurek03}, is due
to extra degrees of freedom (the environment) that interact with the
system, and to imperfections of the gate operations. Overcoming decoherence
is clearly one of the key factors for implementing large scale quantum
computers. Several techniques have been proposed for this purpose,
including dynamical decoupling \cite{5916}, decoherence-free subspaces
\cite{3045}, and quantum error correction \cite{3921,6581}. These
proposals have been tested on small systems of nuclear spins \cite{6013},
or trapped ions \cite{monz:200503} or spin model quantum memories
\cite{biercuk_optimized_2009}.

Tests on larger systems, comprising hundreds or thousands of qubits,
are more difficult. So far, the only physical system that offered
this possibility is nuclear magnetic resonance (NMR) of dipolar coupled
spins \cite{Suter04,Suter06}. Processes that transfer quantum information
over large distances can also be studied in spin chains \cite{Bose2003}.
An example of such a linear spin system was studied by solid-state
NMR \cite{Cappellaro2007}. These model systems do not allow addressing
of individual qubits, but they allow one to study some aspects of
decoherence and information transfer. In particular, they can be used
for studying the effect of the finite precision of experimental quantum
gate operations on the transfer of quantum states: it was predicted
that quantum information cannot be transmitted over arbitrary distances,
but that it will become localized \cite{Anderson1958,Pomeransky2004,Burrell2007,Keating2007,Allcock2009}.

In this paper, we present the first experimental study trying to answer
the following question: How far can quantum information be transmitted
with quantum gate operations of finite precision? For this purpose,
we use an NMR quantum simulator. Starting from individual, uncorrelated
spins, we measure the build-up of clusters of correlated spins of
increasing size. Introducing a perturbation to the Hamiltonian that
generates these clusters, we find, that the size of the clusters reaches
an upper bound. This upper bound appears to be a dynamic equilibrium:
if the cluster size is initially larger than this equilibrium value,
it decreases under the effect of the perturbed Hamiltonian, while
the unperturbed Hamiltonian leads to an increase. The equilibrium
size decreases with increasing strength of the perturbation.

\textit{Growth of spin clusters.---} All the spins of the system are
equivalent and they are in a strong magnetic field. In its Zeeman
rotating-frame, the Hamiltonian of the spin system used for the quantum
simulations is the high-field homonuclear dipolar interaction \cite{Slichter}\begin{align}
 & \widehat{\mathcal{H}}_{dd}=\sum_{i<j}d_{ij}\left[2\hat{I}_{z}^{i}\hat{I}_{z}^{j}-(\hat{I}_{x}^{i}\hat{I}_{x}^{j}+\hat{I}_{y}^{i}\hat{I}_{y}^{j})\right],\label{Hraw}\end{align}
 where $\hat{I}_{x}^{i},\hat{I}_{y}^{i}\mbox{ and }\hat{I}_{z}^{i}$
are spin-1/2 operators and $d_{ij}$ the coupling constants. The quantum
simulations start from the high-temperature thermal equilibrium \cite{Slichter},
$\hat{\rho}_{0}\propto\hat{I}_{z}=\sum\hat{I}_{z}^{i}$. In this state,
the spins are uncorrelated.

\begin{figure}
\includegraphics[bb=96bp 604bp 489bp 761bp,clip,width=7.5cm,height=2.5cm]{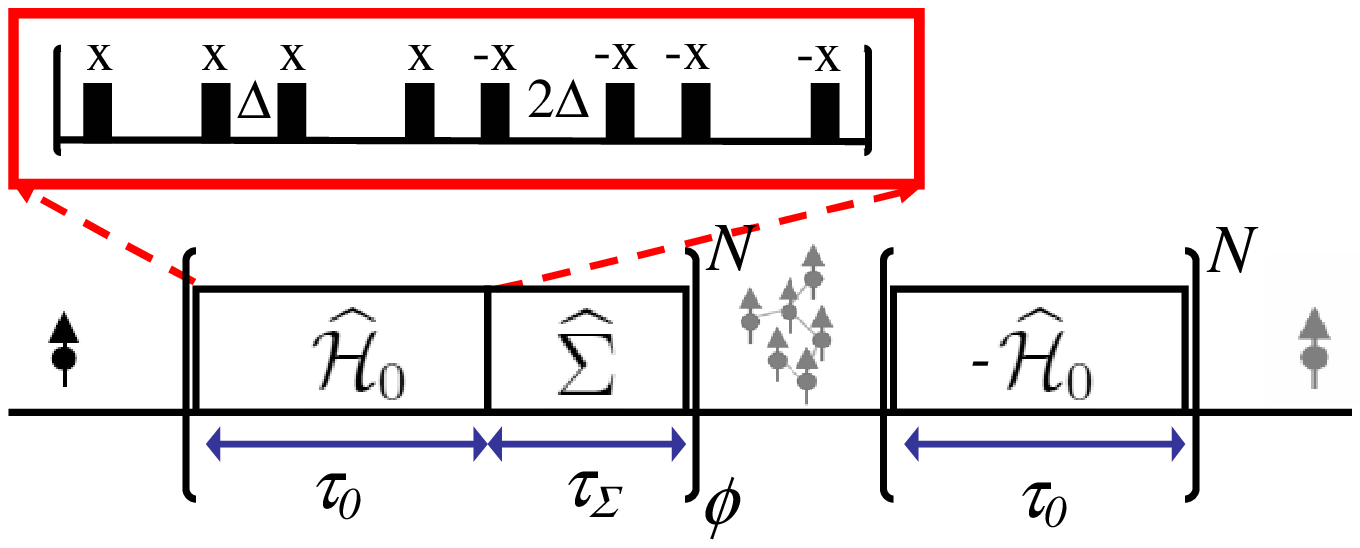}

\caption{(Color online) NMR sequence for the quantum simulations. An unperturbed
evolution is achieved when $\tau_{\Sigma}=0$. The effective Hamiltonian
$\widehat{\mathcal{H}}_{0}$ is generated by the sequence of $\text{\ensuremath{\pi}/2}$
pulses shown in the upper part of the figure.}

\label{fig:Periods} 
\end{figure}

We generate states with correlated spin clusters whose density operator
terms are of the form $\hat{I}_{u}^{i}...\hat{I}_{v}^{j}\hat{I}_{w}^{k}\left(u,v,w=x,y,z\right)$,
by letting the system evolve under the effective Hamiltonian 

\begin{equation}
\widehat{\mathcal{H}}_{0}=-\sum_{i<j}d_{ij}\left[\hat{I}_{x}^{i}\hat{I}_{x}^{j}-\hat{I}_{y}^{i}\hat{I}_{y}^{j}\right].\label{flip-flip}\end{equation}
This Hamiltonian is prepared by means of a standard NMR sequence \cite{5105,Baum1985}
shown in the upper part of Fig. \ref{fig:Periods}. This Hamiltonian
flips simultaneously two spins with the same orientation. Accordingly,
the $z$-component of the magnetization $M_{z}$ changes by $M=\Delta M_{z}=\pm2.$
At the same time, the number $K$ of correlated spins changes by $\Delta K=\pm1$.%

To determine the average number of correlated spins, we use standard
NMR techniques developed by Baum \emph{et al.} \cite{Baum1985}. The
technique relies on the fact that in a system of $K$ spins, the number
of transitions with a given $M$ shows a binomial distribution. For
$K\gg1,$ the binomial distribution can be well approximated with
a Gaussian of width $\propto\sqrt{K}$. To determine the effective
size of the spin clusters in a given state, we decompose its density
operator $\rho$ into components of coherence order $M$. They can
be distinguished experimentally by rotating the system around the
$z-$axis: a rotation $\hat{\phi}_{z}=e^{-i\phi\hat{I}_{z}}$ by $\phi$
changes the density operator to

\begin{equation}
\hat{\rho}\left(\phi\right)=\hat{\phi}_{z}\hat{\rho}\hat{\phi}_{z}^{-1}=\sum_{M}\hat{\rho}_{M}^{{}}e^{iM\phi},\label{eq:rhophi}\end{equation}
 where $\hat{\rho}_{M}$ contains all the elements of the density
operator involving coherences of order $M$. The terms with $M=0$
are zero quantum coherences and populations.

\begin{figure}
\includegraphics[width=8cm,height=6cm]{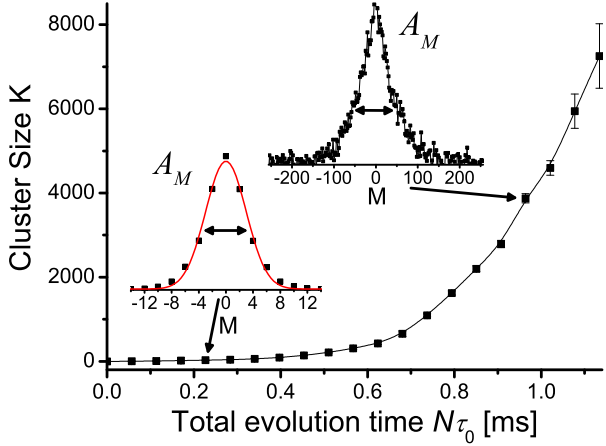}

\caption{(Color online) Time evolution of the cluster size of correlated spins
with the unperturbed Hamiltonian $\widehat{\mathcal{H}}_{0}$. Distributions
of the squared amplitudes $A_{M}$ of density operator components
as a function of the coherence order $M$ are shown for two different
cluster sizes. The latter are obtained from the half-width $2\sqrt{\ln(2K)}$
of the distribution function $A_{M}$.}

\label{fig:Growth} 
\end{figure}

If the system evolves under the Hamiltonian (\ref{flip-flip}), the
cluster size increases indefinitely, as shown in Figure \ref{fig:Growth}.
The figure also shows two examples of $\hat{\rho}_{M}$ distributions.

This evolution can be reversed completely by changing the Hamiltonian
from $\widehat{\mathcal{H}}_{0}$ to $-\widehat{\mathcal{H}}_{0}$.
Experimentally, this is achieved by shifting the phase of all RF pulses
by $\pm\pi/2$ \cite{5105}. This indefinite growth of the cluster
size, as well as the reversibility of the time evolution are no longer
possible if the effective Hamiltonian deviates from the ideal form
(\ref{flip-flip}). This allows us to experimentally induce localization
effects by concatenating short evolution periods under a perturbation
Hamiltonian $\widehat{\Sigma}$ with evolution periods under the ideal
Hamiltonian $\widehat{\mathcal{H}}_{0}$. For the present experiments,
we choose $\widehat{\Sigma}=\widehat{\mathcal{H}}_{dd}$, and we label
the durations of the two time periods $\tau_{\Sigma}$ and $\tau_{0}$,
as shown in Fig. \ref{fig:Periods}. When the duration $\tau_{\mbox{c}}=\tau_{0}+\tau_{\Sigma}$
of each cycle is short compared to the inverse of the dipolar couplings
$d_{ij}$, the resulting evolution can be described by the effective
Hamiltonian\begin{equation}
\widehat{\mathcal{H}}_{\mathrm{eff}}=(1-p)\widehat{\mathcal{H}}_{0}+p\widehat{\Sigma},\label{Heff}\end{equation}
 where the relative strength $p=\tau_{\Sigma}/\tau_{\mbox{c}}$ of
the perturbation can be controlled by adjusting the duration $\tau_{\Sigma}$.
Since the Hamiltonian $\widehat{\mathcal{H}}_{0}$ is generated as
an effective Hamiltonian, it always deviates from the ideal Hamiltonian.
In the experiment, we compare the artificially perturbed evolution
of $\widehat{\mathcal{H}}_{\mathrm{eff}}$ with the $\widehat{\mathcal{H}}_{0}$
evolution with its intrinsec errors. Note that the intrinsic errors
do not produce localization on the time scale of our experiments (see
Fig. \ref{fig:Growth}).

Taking this perturbation into account, and starting from thermal equilibrium,
the state of the system at the end of $N$ cycles is \begin{equation}
\hat{\rho}^{\mathcal{H}_{\mathrm{eff}}}\left(N\tau_{\mbox{c}}\right)=\widehat{U}_{N}^{\dagger}\hat{I}_{z}\widehat{U}_{N},\end{equation}
where $\widehat{U}_{N}=\exp\left\{ -\frac{i}{\hbar}\widehat{\mathcal{H}}_{\mathrm{eff}}N\tau_{\mbox{c}}\right\} $
is the evolution operator for the perturbed evolution. The NMR signal,
which is measured after the backward evolution $\widehat{V}_{N}=\exp\left\{ \frac{i}{\hbar}\widehat{\mathcal{H}}_{0}N\tau_{\mbox{0}}\right\} $,
can be written as $S(N\tau_{\mbox{c}})=\mbox{Tr}\left\{ \mathcal{\widehat{A}}\hat{\rho}^{\mathcal{H}_{\mathrm{eff}}}\left(N\tau_{\mbox{c}}\right)\right\} $,
where \begin{equation}
\mathcal{\widehat{A}}=\widehat{V}_{N}\hat{I}_{z}\widehat{V}_{N}^{\dagger}=\hat{\rho}^{\mathcal{H}_{0}}\left(N\tau_{\mbox{0}}\right)\end{equation}
 is the effective observable and $\hat{\rho}^{\mathcal{H}_{0}}$ the
density operator of the unperturbed evolution. We again determine
the cluster size by applying rotations $\hat{\phi}_{z}$ around the
$z$-axis, as in Eq. (\ref{eq:rhophi}). The resulting NMR signal
is then

\begin{multline}
S\left(\phi,N\tau_{\mbox{c}}\right)=\sum_{M}\mbox{\ensuremath{e^{i\phi M}}}A_{M}\\
=\sum_{M}\mbox{\ensuremath{e^{i\phi M}}Tr}\left\{ \hat{\rho}_{M}^{\mathcal{H}_{0}}\left(N\tau_{\mbox{0}}\right)\hat{\rho}_{M}^{\mathcal{H}_{\mathrm{eff}}}\left(N\tau_{\mbox{c}}\right)\right\} .\end{multline}

For ideal evolution ($p=0$), the individual terms $A_{M}$ in the
last equation correspond to the squared amplitudes of density operator
elements $\hat{\rho}_{M}^{\mathcal{H}_{0}}\left(N\tau_{\mbox{0}}\right)$
with coherence order $M$. For perturbed evolution, $(p\ne0)$, they
are reduced by the overlap of the actual density operator elements
$\hat{\rho}_{M}^{\mathcal{H}_{\mathrm{eff}}}\left(N\tau_{\mbox{c}}\right)$
with the ideal ones. To extract these amplitudes from the experimental
data, we perform a Fourier transform with respect to $\phi$. Two
examples for the resulting $A_{M}$ are shown in the insets of Fig.
\ref{fig:Growth}.

\begin{figure}
\includegraphics[clip,width=8cm,height=8cm]{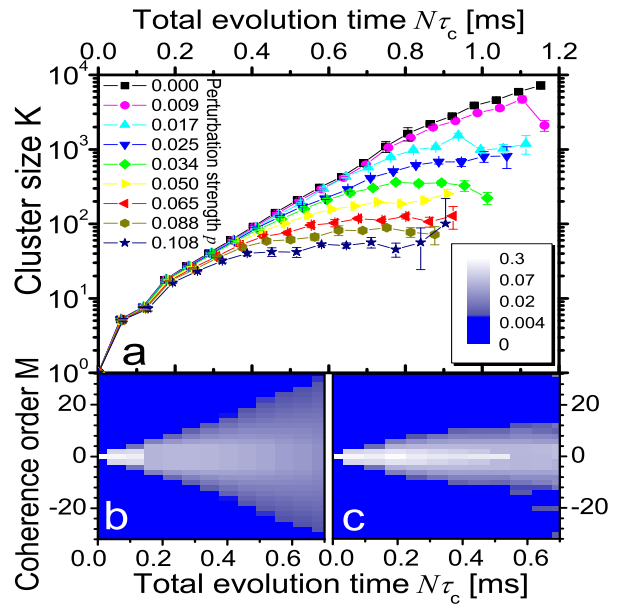}

\caption{(Color online) a) Time evolution of the cluster size. The black squares
represent the unperturbed time evolution and the other symbols correspond
to different perturbation strengths according to the legend. b,c)
Distributions of the amplitudes $A_{M}\left(N\tau_{\mbox{c}}\right)$
for unperturbed dynamics (b; $p=0$) and a perturbed evolution (c;
$p=0.108$) respectively. The perturbed evolution in panel c shows
localization at a cluster size $K_{\mbox{loc}}$$\simeq$56 spins.}

\label{fig:Kvst} 
\end{figure}

\textit{Experimental results.---} Experiments were performed on a
home-built solid state NMR spectrometer with a $^{\text{1}}$H resonance
frequency of 300 MHz. The spins are the protons of polycrystalline
adamantane where the strength of the dipolar interaction, quantified
by the second moment of the resonance line is $7.9$ kHz. In the experiments
we chose $\tau_{0}=57.6\mu\mbox{s}$. The black squares of Fig. \ref{fig:Kvst}a
shows the averaged number of correlated spins as a function of time
for an unperturbed evolution, $p=0$. The observed cluster size $K(N\tau_{\mbox{c}})$
grows almost exponentially over the range considered here \cite{Lacelle1991}.
The other symbols of panel (a) show the evolution of the number of
correlated spins for different values of $p$. Initially, the cluster
size $K(N\tau_{\mbox{c}})$ starts to grow as in the unperturbed evolution,
but then it saturates after a time that decreases with increasing
perturbation strength $p$. We consider this as evidence of localization
due to the perturbation. The size of the cluster at which this saturation
occurs is also determined by the strength of the perturbation: increasing
perturbation strength reduces the limiting cluster size. Panels b
and c of Fig. \ref{fig:Kvst} visualize this localization directly
by comparing the generation of high-order multiple quantum coherences
for unperturbed (panel b) and perturbed (panel c; $p=0.108$) evolution:
they give a color-coded representation of the amplitudes $A_{M}\left(N\tau_{\mbox{c}}\right)$
as a function of evolution time $N\tau_{\mbox{c}}$. While the distribution
spreads continuously in panel b, it reaches a limiting value in panel
c.

\begin{figure}
\includegraphics[width=7.5cm,height=8cm]{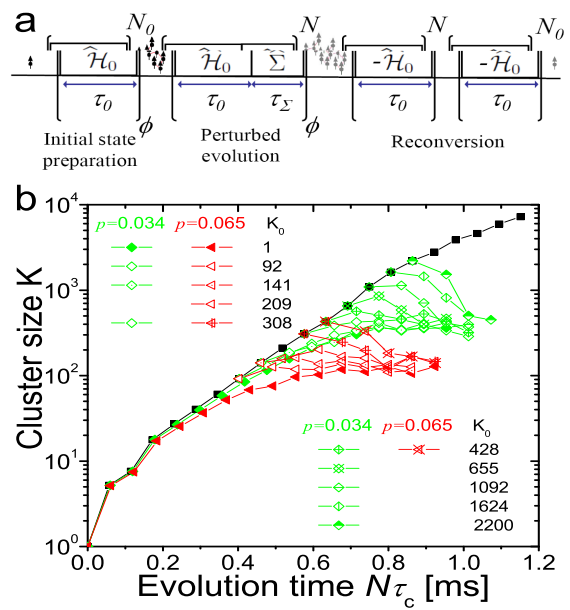}

\caption{(Color online) a) NMR pulse sequence for preparing different initial
clusters sizes and subsequently evolving them in the presence of a
perturbation. b) Time evolution of the correlated cluster size starting
from different initial sates. Filled symbols are evolutions from an
uncorrelated initial state for two different perturbation strengths
given in the legend. Empty symbols start from an initial state with
$K_{0}$ correlated spins.}

\label{fig:localization_convergence} 
\end{figure}

While these experiments show that the cluster size reaches a stationary
value, they leave open the question if this limiting size results
from a slow-down in the growth \cite{Burrell2007} or it represents
a dynamic equilibrium state. We therefore repeated the above experiment
for a series of initial conditions corresponding to different clusters
sizes. Figure \ref{fig:localization_convergence}a shows the corresponding
pulse sequence: The initial state preparation, consisting of an evolution
of duration $N_{0}\tau_{0}$ under the unperturbed Hamiltonian $\widehat{\mathcal{H}}_{0}$,
generates clusters of size $K_{0}$. During the subsequent perturbed
evolution of duration $N\tau_{\mbox{c}}$, these initial clusters
grow or shrink. Figure \ref{fig:localization_convergence}b shows
the results for two perturbation strengths, $p=0.034$ and $p=0.065$.
The filled symbols correspond to uncorrelated initial states and the
empty symbols to various initial cluster sizes $K_{0}$. The experimental
results clearly show that, for a given perturbation strength, the
size of the spin clusters tends towards the same limiting value, independent
of the initial condition. We verified this behavior for additional
perturbation strengths (data not shown in the figure).

\begin{figure}
\includegraphics[bb=58bp 21bp 718bp 522bp,scale=0.3]{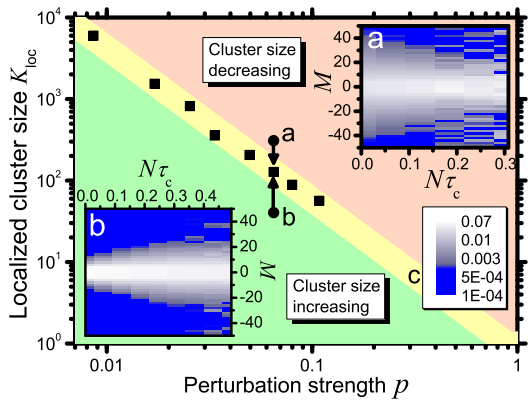}

\caption{(Color online) Localized cluster size $K_{\mbox{loc}}$ (square symbols)
of correlated spins versus the perturbation strength $p$. Three dynamical
regimes for the evolution of the cluster size are identified depending
of the number of correlated spins compared with the perturbation dependent
localization value: a) a cluster size decreases, b) a cluster size
increases c) stationary regime.}

\label{fig:localizednumber} 
\end{figure}

Figures \ref{fig:Kvst}a and \ref{fig:localization_convergence}b
indicate that the size of the resulting clusters decreases with increasing
strength of the perturbation. To establish this dependence in a quantitative
manner, we determined the size of the localized clusters from the
data shown in Fig. \ref{fig:Kvst} and plotted them against the perturbation
strength (black squares in Fig. \ref{fig:localizednumber}). The diagonal
line in Figure \ref{fig:localizednumber} represents a linear fit
to the experimental data represented by the black squares; its width
indicates the error of the fit. A functional dependence $K_{\mbox{loc}}\sim p^{-1.86\pm0.05}$
is obtained, indicating that the size of the localized clusters decreases
with the square of the perturbation strength. The limiting value for
$p=1$ is $K_{\mbox{loc}}\approx1$, indicating that the system becomes
completely localized if the perturbation strength is significantly
larger than the unperturbed Hamiltonian. The figure also summarizes
the evolution of the cluster size before the static (localized) size
is reached: If the initial size is larger than the stationary value
for the given perturbation strength, $K_{0}>K_{\mbox{loc}}$, the
cluster shrinks (inset a in the figure, above the diagonal). If it
is smaller, $K_{0}<K_{\mbox{loc}}$, the size increases (inset b,
below the diagonal).

\textit{Discussion and Conclusions.---} Decoherence has long been
recognized to limit the time for which quantum information can be
used. Here, we have shown that it also limits the distance over which
quantum information can be transferred. To demonstrate this effect,
we have compared the spreading of information in a system of nuclear
spins under the influence of a Hamiltonian that transfers information
and a perturbation Hamiltonian of variable strength. In combination,
these opposing forces result in a quantum state that becomes localized.
The localization size decreases with increasing strength of the perturbation.
Our experimental result of a dynamic equilibrium size of the localized
state differs from theoretical predictions that only indicate a slow
down of the spreading \cite{Burrell2007}. The experiments were performed
with nuclear spins, which we use as a quantum simulator, and the perturbation
is taken as a model for the disorder considered in the discussion
of localization \cite{Anderson1958,Pomeransky2004,Burrell2007,Keating2007,Allcock2009}.

These results may also be connected to our earlier findings that the
decoherence rate of quantum states with many correlated qubits increases
with the size of the system \cite{Suter04}, indicating that larger
systems are more sensitive to perturbations. As the system size increases,
the tendency for the system to spread is therefore balanced by the
restriction due to the perturbation. As a heuristic argument, we note
that in a suitable interaction representation, the perturbation will
cause a decay whose rate may be calculated by second order perturbation
theory. We expect there a quadratic dependence on the perturbation
strength that could be the source of the dynamic equilibrium size
behaviour. The results presented here provide information about the
spatial bounds for transferring quantum information in a spin network
and indicate how precise manipulations of large quantum systems have
to be. 
\begin{acknowledgments}
Acknowledgments.--- GAA thanks the Alexander von Humboldt Foundation
for a Research Scientist Fellowship. We thank Marko Lovric, Hans Georg
Krojanski and Ingo Niemeyer for helpful discussions and technical
support. 
\end{acknowledgments}
\bibliographystyle{apsrev} \bibliographystyle{apsrev}
\bibliography{bibliography,BibsDS}

\end{document}